**Multistable polar textures in geometrically frustrated nematic liquid crystals**

Ufuoma I. Kara[1,#], Boyuan Chen[1,#], Simon Čopar[2,#], Shucong Li[3], Rajdeep Mamtani[1], Xu Yang[1], Eric Boerner[1], Zhan Yang[1], Alan H. Weible[1], Yuxing Yao[4], Robin L. B. Selinger[5,6], Uroš Tkalec[7,8,9],*, Xiaoguang Wang[1,10],*

[1]William G. Lowrie Department of Chemical and Biomolecular Engineering, The Ohio State University, Columbus, OH, 43210, USA

[2]Faculty of Mathematics and Physics, University of Ljubljana, Jadranska cesta 19, 1000 Ljubljana, Slovenia

[3]Department of Chemistry and Chemical Biology, Harvard University, Cambridge, MA, 02138, USA

[4]Division of Chemistry and Chemical Engineering, California Institute of Technology, Pasadena, CA, 91125, USA

[5]Advanced Materials and Liquid Crystal Institute, Kent State University, Kent, OH, 44242, USA

[6]Department of Physics, Kent State University, Kent, OH, 44242, USA

[7]Institute of Biophysics, Faculty of Medicine, University of Ljubljana, Vrazov trg 2, 1000 Ljubljana, Slovenia

[8]Faculty of Natural Sciences and Mathematics, University of Maribor, Koroška cesta 160, 2000 Maribor, Slovenia

[9]Department of Condensed Matter Physics, Jožef Stefan Institute, Jamova cesta 39, 1000 Ljubljana, Slovenia

[10]Sustainability Institute, The Ohio State University, Columbus, OH, 43210, USA

*Corresponding authors: U.T. (uros.tkalec@mf.uni-lj.si), X.W. (wang.12206@osu.edu)

#These authors contributed equally to this work.

**Abstract**

The ability to manipulate polar entities with multiple external fields opens exciting possibilities for emerging functionalities and novel applications in spin systems, photonics, metamaterials, and soft matter. Liquid crystals (LCs), exhibiting both a crystalline structure and liquid fluidity, represent a promising platform for manipulating phases with polar molecular order, notably ferroelectric ones. However, achieving a polar symmetry is challenging with rod-shaped LC molecules, which form predominantly apolar nematic phases. We report an approach in which a geometric lattice confinement of nematic LCs is used to induce planar polar order on the scale of a mesoscopic metamaterial. We confine the nematic LC in a micropillar array, forming topological defect−pillar pairs of elastic dipoles with a free top interface in contact with an immiscible fluid. The resulting dipole lattice configurations can be programmed rheologically by flowing the top fluid and maintained even after flow cessation, a phenomenon attributed to orientational multistability of the dipoles. This multi-memory effect enables the encoding and reconfiguration of directional information. Overall, this research establishes a foundational understanding of topological dipoles under confinement and shear flow, enabling the detection, tracking, and recording of flow profiles and paving the way for future advances in soft matter physics and stimuli-responsive materials.

**Introduction**

Polar order, a ubiquitous and fundamental property across diverse scientific domains, arises from the asymmetric distribution of charges, states, or orientations within a system. This asymmetry leads to the formation of two distinct and opposing entities and consequent dipole−dipole coupling, crucial for understanding and predicting complex interactions and behaviors of matter at different scales, from subatomic particles to macroscopic structures[1]. The organized arrangement of spins in oxide nanostructures, (anti)ferroelectric, and (anti)ferromagnetic materials imparts unique electric, magnetic and optical properties[2-5]. Extending this foundational concept, frustrated artificial spin ice systems demonstrate how manipulating spin interactions can unveil emergent multipole arrangements[6,7], offering new avenues in magneto−optical properties and information storage[8]. Similarly, mechanical metamaterials exploit dipole-like responses inherent in topological states, such as auxetic structures and Kagome lattices, to achieve customizable mechanical properties[9-11]. The precise assembly of photonic spin lattices in metasurfaces facilitates unparalleled control over light polarization, opening up possibilities in optical communication and sensing[12-15]. Thus, the control and reconfiguration of polar textures enable designing advanced electronics, photonics, and energy technologies and the exploration of novel physical phenomena.

The intricate phenomenology of liquid crystals (LCs) combines the ordered structure of crystalline solids with the inherent fluidity of liquids[16]. Rod-shaped molecules typically form LC phases with apolar orientational order, such as the extensively studied nematic phase[17,18]. Lately, bent-core and conical-shaped

LC molecules have emerged, naturally exhibiting (anti)ferroelectric alignment and polar phases with inherent permanent electric dipole moments and ferroelectric properties[19-24]. Despite advancements in achieving polar nematic phases in complex rod-shaped LC molecules like DIO, RM734 and nBOE[25-27], molecular structure design requires multiple dipoles with large net dipole moments, significantly restricting polar order to specific compounds. At the mesoscale, establishing directionality without inherent head−tail symmetry necessitates techniques such as confinement[28-33], inclusions[34-38], electric field[39,40], and flow of active[41-45] and passive fluids[46-49]. These methods disrupt the nematic symmetry, resulting in the formation of topological defects that induce dipolar order within the LC director profile, a phenomenon especially prominent at the micrometer scale. Notable examples include the formation of elastic dipoles between colloidal microspheres and associated hyperbolic hedgehog defects[34-38], as well as the emergence of polar order in thin slab microfluidics through flow-driven transitions[46-49]. However, the systems are primarily constrained to one-dimensional orientations due to preferred directions imposed by external means like surface rubbing or flow direction. Creating a true two-dimensional (2D) system based on apolar nematic LCs, where free dipole rotation within a well-organized lattice enables significant dipole−dipole interactions among neighbours, akin to those in spin ice and photonic crystals, remains an open challenge.

Here, we report a pioneering approach to induce polar order and dipole lattice in simple rod-shaped LCs through geometric confinement. By utilizing homeotropically treated silicon (Si) micropillar arrays as geometric boundaries, we generate topological defects analogous to those observed around spherical colloidal inclusions, but with the advantage of a pre-defined fixed lattice. This system results in the formation of a lattice of topological elastic dipoles, characterized by surface boojum topological defects with a winding number of −1 adjacent to each micropillar. The interactions and orientation of neighbouring dipoles, intricately linked to the pillar array, superimpose polar order at the lattice scale onto the molecular orientational order profile within the confined nematic LCs. While mimicking a 2D crystal lattice in solid systems, the associated defects remain mobile, enabling structural variations such as domain walls, vacancies, and instances of double occupancy. Remarkably, our experiments demonstrate that elastic dipoles reorient under shear induced by a flowing fluid, enabling the storage and active modulation of directional information inherent to these dipoles. Overall, our design of a multistable, reconfigurable dipole lattice in apolar nematic LCs enables short-range dipolar interaction and dynamic complexities observed in conventional polar systems, opening new avenues for designing advanced soft materials with programmable functionalities.

**Results**

We infused LCs in homeotropically treated Si micropillar arrays (see Methods), focusing specifically on rod-shaped LC molecules that exclusively exhibit an apolar nematic phase, such as 4'-pentyl-4-biphenylcarbonitrile (5CB) and E7, a nematic mixture of cyanobiphenyl and cyanoterphenol components. Unless stated otherwise, the Si micropillar arrays had a hexagonal arrangement with dimensions of 40 μm in diameter, 30 μm in height, and a pitch of 70 μm (defined as the center-to-center distance between two nearest micropillars). When 5CB was introduced into homeotropically treated micropillar arrays, we observed a symmetric optical appearance and the emergence of a topological defect ring with a winding number of −1/2 around each micropillar under a polarized light microscope in reflection mode (Figs. 1a-1c). This LC orientation results from the homeotropic ordering of LCs at the air−LC interface (Fig. 1b), which is consistent with previous study[30]. In contrast, untreated Si micropillar arrays exhibited uncontrolled defect structures (Supplementary Fig. 1).

Subsequently, we introduced an immiscible water film into contact with the 5CB-infused micropillar array, disrupting the inherent symmetry at the boundary and inducing a transition from perpendicular alignment at the air−LC interface to degenerate planar anchoring at the water−LC interface (Fig. 1d). The cross-section, with a radial profile as shown in Fig. 1e, illustrates the shift from a ring defect to a surface boojum with a winding number of −1 in the top projection. The micropillar array-confined LCs act as a 2D nematic system featuring only whole winding numbers of point defects. Fig. 1e reveals an association between each surface boojum and an individual micropillar. We refer to these paired arrangements as topological elastic dipoles, with directions defined from the defect to the pillar, resembling colloidal inclusion−defect dipole configurations reported in previous studies[34-38]. Furthermore, the dipoles tend to orient towards the neighbouring micropillars, endowing our system with properties akin to a lattice of dipoles that can assume six possible orientations (Supplementary Movie 1), evoking characteristic dipole−dipole neighbour interactions in planar spin systems. In addition, deviations from this model are observed, as some point defects may be displaced from their lattice position, resulting in vacancies (micropillars without an associated defect) and double occupancy (two point defects surrounding a single micropillar), as shown in Fig. 1f and Supplementary Fig. 2. These displaced point defects can be analogized to electron−hole pairs in semiconductors.

Expanding upon these insights, we examined the micropillar array's geometrical characteristics—pitch and lattice structure—and their effects on LC ordering and elastic dipole configurations (Supplementary Figs. 3 and 4 and Note 1). We observed that at a 50 μm pitch, pairs of vertical defect lines with −1/2 winding number emerged around micropillars, contrasting with the single −1 point defect observed at a 70 μm pitch. Increasing the pitch to 100 μm diminished dipole interactions, weakening dipole−dipole coupling within the lattice. Moreover, employing a square lattice configuration, in contrast to a hexagonal one, results in a reduction of stable dipole orientations from six to four. These observations demonstrate the profound impact of geometrical confinement on dipole interaction and orientation.

By treating the LC orientational ordering profile as a unit vector field, we assigned signature (depicted by the direction of arrows) to micropillars (distinct from their +1 winding number). Identical signatures between neighbouring pillars necessitated the presence of π solitons. In contrast, opposite signatures allowed for a uniform director alignment (inset in Fig. 1g). Solitons involve more deformation, resulting in higher elastic energy in lattices of micropillars with identical signatures. However, orientation by flow-alignment or directional passage of a phase transition front during cooling acts on all micropillar surroundings in a coherent manner, favoring parallel dipole alignment and equal signatures. Once formed, the relative signatures of the micropillars are topologically protected and remain unchanged during manipulations despite being in a higher energy state[50]. The resulting solitons appear as straight lines in a uniformly aligned lattice and branch or terminate in complex orientations of dipoles. While there could be combinatorial myriads of possible ±1 signature states in this system, our experiments consistently demonstrate that most micropillars exhibit the same polarity, greatly restricting the number of accessible states.

A numerical sketch in Fig. 1g illustrates the above phenomenon, showing predominantly same-polarity states on the left half and mixed polarities on the right half. This observation arises because the imposition of polar order always progresses as a moving front. The resulting dipole reconfiguration consistently produces states with numerous solitons and, consequently, higher elastic energy, which is challenging to achieve with alternative methods. The capability to create perfectly uniform structures is crucial for all experimental outcomes discussed in this study. Notably, the system's local metastability, which arises from the presence of six possible dipole orientations, combined with the topological protection of signatures, fosters a switchable multistable system rather than one predominantly governed by energy. This analysis simplifies the description and control, as the remaining consideration is the relative orientation of neighbouring elastic dipoles, along with occasional lattice vacancies and dislocations. Consequently, our system can be effectively modeled as an XY system with nearest-neighbour interaction, proportional to the squared difference of orientation angles, favoring parallel dipole alignment, along with an additional potential with minima at multiples of 60º to ensure metastability (see Methods for details and Supplementary Movie 2 for a qualitative match with experimental relaxation). Our simulations of the XY system demonstrate that a model employing simplified nearest-neighbour interactions captures the essential dynamics effectively and accurately delineates the fundamental dipole reorientation processes. However, the emergence of complex phenomena, such as isolated pillars of opposing polarities, dislocated defects, and double occupancy, underscores the limitations of this simplified model. These intricate features are, nonetheless, comprehensively understood within the framework of soliton and defect topology, as elucidated in prior discussions.

Next, we aimed to manipulate the dipole arrangement experimentally using surface shear. Initially, a 10 μL water droplet was placed onto the LC surface, and we observed the dipole orientation around the droplet's center. The observations, as depicted in Figs. 2a and 2b, revealed the dipoles radially oriented away from the central micropillar, except for the central micropillar, which displayed an opposite signature, with the director pointing radially straight to the neighbouring pillars and no accompanying defect (a vacancy). Upon subsequent withdrawal of 5 μL of the water droplet, we observed either a similar radial arrangement with the dipoles pointing toward the central vacancy (Figs. 2c and 2d) or the generation of a mobile monopole (+1 point defect; Fig. 2e and Supplementary Movie 3). The annihilation of this monopole with a −1 defect associated with a micropillar resulted in the creation of a vacancy (the net charge of the pillar is +1; Fig. 2f and Supplementary Movie 4). Since the pillar before annihilation had the same signature as the others, the resulting vacancy did not exhibit the same radial structure as shown in Figs. 2a and 2b. Instead, we observed a spiral structure, which is necessary to ensure the topologically protected additional π turn due to opposite parity of the central vacancy. The arrangement around the central pillar follows the predicted logarithmic spiral structure found in the well-known "magic spiral" solution in a 2D nematic system[16,17,50].

To achieve uniform orientation of the elastic dipoles, we applied a unidirectional flow. As a water droplet slid unidirectionally toward the 6 o'clock position, the dipole states beneath the droplet transformed into a monodomain, with all dipoles aligning parallel to the direction of the water droplet's motion (Fig. 3a and Supplementary Movie 5). This behavior was consistently observed when the water droplet slid in other directions within the hexagonal lattice. Sliding the droplet in intermediate directions caused the dipoles to align with the nearest stable state. When the intermediate direction aligned with the angle between two stable states, we observed an equal probability of the dipoles orienting toward either stable state (Supplementary Fig. 5 and Note 2). Introducing an air bubble sliding underwater further illustrated the capacity to achieve multiple domains of dipole alignment (Supplementary Fig. 6 and Note 3). Notably, the dipoles retained their orientation even after the motion of the droplet or the bubble stopped.

The above phenomenon implies a characteristic multi-memory effect of micropillar array-confined LCs, leveraging the synergistic impact of shear-induced alignment and temperature-induced randomization of dipoles. As shown in Fig. 3a, we encoded directional information onto the LC surface by sliding droplets on the LC surface. This encoded information can be efficiently erased by inducing a phase transition to the isotropic phase. Furthermore, the encoded information can be reconfigured by re-sliding the droplet in a different direction. The reorientation requires migration of defects, which is local, and rearrangement or even reversal of solitons between the dipoles, which involves a global reconfiguration of the director (Fig. 3a, bottom row). Remarkably, the encoded information remains stable for at least 300 days using a glycerol droplet, as evidenced in Supplementary Fig. 7. The encoding process, which includes erase, rewrite and overwrite, can be repeated for at least 1,000 cycles without any loss of function (Supplementary Fig. 8). In addition, the shear force for dipole reorientation is minimal, at approximately 2.4 pN per dipole, highlighting the intrinsic sensitivity of apolar nematic LCs (Supplementary Fig. 9, Note 4 and Movie 6). These results suggest that rheological manipulation can effectively control dipole orientation, which can complement traditional optical, acoustic, and electric field techniques[36-40].

In contrast to the behavior observed with a single droplet, which typically generates a monodomain, merging two droplets introduces more complex dipole domain patterns, separated by either a soliton, Ising, or Néel domain walls, demonstrating the capability to encode diverse domain configurations. This situation is analogous to the superposition of dual external fields on a dipole lattice (Fig. 3b). Furthermore, our experiments with underwater streaming demonstrated the capability to create complex domain patterns without the domain walls typically formed during droplet merging, accessing dipole states not possible with droplet merging alone, as detailed in Supplementary Fig. 10.

In the final set of experiments, we aimed for an enhancement in the dipole orientation control resolution using a micro-cantilever, as schemed in Fig. 4a. Fig. 4b and Supplementary Movie 7 demonstrate the efficacy and precision of this method to manipulate a single dipole. Figs. 4c and 4d and Supplementary Fig. 11 and Movie 7 further extend our capability to manipulate arrays of dipoles, creating single and intersecting lines and complex patterns, including a radial pattern and a 'smiley face.' Overall, these findings demonstrate the synergy of external actuation and geometrical confinement in encoding complex, reconfigurable patterns in the elastic dipole lattice.

**Discussion**

In summary, we report an innovative use of geometrical confinement, particularly with hexagonally arranged micropillar arrays, to induce polar order in traditionally apolar nematic phases of simple rod-shaped LC molecules. The disruption of rotational symmetry at the interfaces of confined LCs results in a quantized 6-fold lattice of elastic dipoles superimposed on the apolar nematic director field. This coupling between the long-range orientational order of the nematic phase and the short-range arrangement of elastic dipoles creates an artificial nematic multiferroic system with controllable "magnetization" and elastic field. The confined LC system exhibits diverse spin-like patterns, including a spin glass state with random dipole orientations influenced by pillar winding numbers, allowing for structural deviations like vacancies and double occupancy. Under external fields, the system manifests various patterns, including nontrivial hedgehogs, chiral (anti)vortices, and different point defects with distinct elastic dipole ordering, along with displaying uniform, stripe, labyrinthine, and polycrystalline domains of collectively aligned dipoles, separated by diverse domain walls such as solitons, Ising walls, and Néel walls. Moreover, the confined LC system demonstrates topological analogies with other condensed matter systems, offering potential applications in reconfigurable materials such as fluidic circuits. The characteristic coupling between neighbouring dipoles and the multi-memory effect of microstructure-confined LCs provides a versatile platform for exploring fundamental principles and phenomena in dipole lattice structures.

While our focus has been primarily on hexagonal and square lattice geometry, future investigations will explore Kagome arrays and the influence of LC mesophases on topological dipole lattices is under exploration. In addition, our findings highlight the utility of flow alignment as a straightforward, adaptable method for orientational control. Our future efforts will aim to explore its promising applications as a flow sensor within microfluidics and active or biological systems exhibiting self-driven motion. Last, we will explore defect structure in geometrically confined LCs of other LC mesophases, including the smectic, cholesteric, and blue phases. On the basis of past studies on ferroelectric properties and supramolecular-scale polarization modulation in smectic LCs[19,20], we anticipate that geometric confinement, along with flow alignment, will unlock novel strategies for manipulating defect structures and enhancing the dynamic properties of these LC systems.

**References**

1. Eerenstein, W., Mathur, N. D. & Scott, J. F. Multiferroic and magnetoelectric materials. *Nature* **442**, 759–765 (2006).
2. Hellman, F. et al. Interface-induced phenomena in magnetism. *Rev. Mod. Phys.* **89**, 025006 (2017).
3. Spaldin, N. A. & Ramesh, R. Advances in magnetoelectric multiferroics. *Nat. Mater.* **18**, 203–212 (2019).
4. Nataf, G. F. et al. Domain-wall engineering and topological defects in ferroelectric and ferroelastic materials. *Nat. Rev. Phys.* **2**, 634–648 (2020).
5. Junquera, J. et al. Topological phases in polar oxide nanostructures. *Rev. Mod. Phys.* **95**, 025001 (2023).
6. Nisoli, C., Moessner, R. & Schiffer, P. Artificial spin ice: Designing and imaging magnetic frustration. *Rev. Mod. Phys.* **85**, 1473–1490 (2013).
7. Skjærvø, S. H. et al. Advances in artificial spin ice. *Nat. Rev. Phys.* **2**, 13–28 (2020).
8. Keim, N. C. et al. Memory formation in matter. *Rev. Mod. Phys.* **91**, 035002 (2019).
9. Bertoldi, K., Vitelli, V., Christensen, J. & van Hecke, M. Flexible mechanical metamaterials. *Nat. Rev. Mater.* **2**, 17066 (2017).
10. Chen, T., Pauly, M. & Reis, P. M. A reprogrammable mechanical metamaterial with stable memory. *Nature* **589**, 386–390 (2021).
11. Xia, X., Spadaccini, C. M. & Greer, J. R. Responsive materials architecture in space and time. *Nat. Rev. Mater.* **7**, 683-701 (2022).
12. Fadeyeva, T. A. et al. Spatially engineered polarization states and optical vortices in uniaxial crystals. *Opt. Express* **18**, 10848-10863 (2010).
13. El Ketara, M., Kobayashi, H. & Brasselet, E. Sensitive vectorial optomechanical footprint of light in soft condensed matter. *Nat. Photon.* **15**, 121-124 (2021).
14. Khanikaev, A. B. & Shvets, G. Two-dimensional topological photonics. *Nat. Photon.* **11**, 763–773 (2017).
15. Solntsev, A. S., Agarwal, G. S. & Kivshar, Y. S. Metasurfaces for quantum photonics. *Nat. Photon.* **15**, 327–336 (2021).
16. de Gennes, P.-G. & Prost, J. *The Physics of Liquid Crystals*. Oxford University Press (1993).
17. Kleman, M. & Lavrentovich, O. D. *Soft Matter Physics: An Introduction.* Springer-Verlag (2003).
18. Bukusoglu, E., Bedolla Pantoja, M., Mushenheim, P. C., Wang, X. & Abbott, N. L. Design of responsive and active (soft) materials using liquid crystals. *Annu. Rev. Chem. Biomol. Eng.* **7**, 163–196 (2016).
19. Meyer, R. B., Liebert, L., Strzelecki, L. & Keller, P. Ferroelectric liquid crystals. *J. Phys. Lett.* **36**, L69–L71 (1975).
20. Coleman, D. A. et al. Polarization-modulated smectic liquid crystal phases. *Science* **301**, 1204–1211 (2003).
21. Basnet, B. et al. Soliton walls paired by polar surface interactions in a ferroelectric nematic liquid

crystal. *Nat. Commun.* **13**, 3932 (2022).
22. Sebastián, N. et al. Polarization patterning in ferroelectric nematic liquids via flexoelectric coupling. *Nat. Commun.* **14**, 3029 (2023).
23. Caimi, F. et al. Fluid superscreening and polarization following in confined ferroelectric nematics. *Nat. Phys.* **19**, 1658–1666 (2023).
24. Yang, J., Zou, Y., Li, J., Huang, M. & Aya, S. Flexoelectricity-driven toroidal polar topology in liquid-matter helielectrics. *Nat. Phys.* doi.org/10.1038/s41567-024-02439-7 (2024).
25. Mandle, R. J. A new order of liquids: polar order in nematic liquid crystals. *Soft Matter* **18**, 5014 (2022).
26. Sebastián, N., Čopič, M. & Mertelj, A. Ferroelectric nematic liquid-crystalline phases. *Phys. Rev. E* **106**, 021001 (2022).
27. Nishikawa, H. et al. Emergent ferroelectric nematic and heliconical ferroelectric nematic states in an achiral "straight" polar rod mesogen. *Adv. Sci.* **11**, 202405718 (2024).
28. Kim, J. H., Yoneya, M. & Yokoyama, H. Tristable nematic liquid crystal device using micropatterned surface alignment. *Nature* **420**, 159–162 (2002).
29. Murray, B. S., Pelcovits, R. A. & Rosenblatt, C. Creating arbitrary arrays of two-dimensional topological defects. *Phys. Rev. E* **90**, 052501 (2014).
30. Araki, T., Buscaglia, M., Bellini, T. & Tanaka, H. Memory and topological frustration in nematic liquid crystals confined in porous materials. *Nat. Mater.* **10**, 303–309 (2011).
31. Cavallaro Jr, M. et al. Exploiting imperfections in the bulk to direct assembly of surface colloids. *Proc. Natl. Acad. Sci. USA* **110**, 18804–18808 (2013).
32. Kim, D. S., Čopar, S., Tkalec, U. & Yoon, D. K. Mosaics of topological defects in micropatterned liquid crystal textures. *Sci. Adv.* **4**, eaau8064 (2018).
33. Wang, X. et al. Moiré effect enables versatile design of topological defects in nematic liquid crystals. *Nat. Commun.* **15**, 497 (2024).
34. Poulin, P., Stark, H., Lubensky, T. C. & Weitz, D. A. Novel colloidal interactions in anisotropic fluids. *Science* **275**, 1770–1773 (1997).
35. Stark, H. Physics of colloidal dispersions in nematic liquid crystals. *Phys. Rep.* **351**, 387-474 (2001).
36. Muševič, I., Škarabot, M., Tkalec, U., Ravnik, M. & Žumer, S. Two-dimensional nematic colloidal crystals self-assembled by topological defects. *Science* **313**, 954–958 (2006).
37. Ognysta, U. et al. 2D interactions and binary crystals of dipolar and quadrupolar nematic colloids. *Phys. Rev. Lett.* **100**, 217803 (2008).
38. Tkalec, U. & Muševič, I. Topology of nematic liquid crystal colloids confined to two dimensions. *Soft Matter* **9**, 8140–8150 (2013).
39. Sasaki, Y. et al. Large-scale self-organization of reconfigurable topological defect networks in nematic liquid crystals. *Nat. Commun.* **7**, 13238 (2016).
40. Migara, L. K. & Song, J.-K. Standing wave-mediated molecular reorientation and spontaneous

formation of tunable, concentric defect arrays in liquid crystal cells. *NPG Asia Mater.* **10**, e459 (2018).
41. Needleman, D. & Dogic, Z. Active matter at the interface between materials science and cell biology. *Nat. Rev. Mater.* **2**, 1–14 (2017).
42. Zhang, R. et al. Spatiotemporal control of liquid crystal structure and dynamics through activity patterning. *Nat. Mater.* **20**, 875–882 (2021).
43. Zhang, R., Mozaffari, A. & de Pablo, J. J. Autonomous materials systems from active liquid crystals. *Nat. Rev. Mater.* **6**, 437–453 (2021).
44. Shankar, S., Souslov, A., Bowick, M. J., Marchetti, C. & Vitelli, V. Topological active matter. *Nat. Rev. Phys.* **4**, 380–398 (2022).
45. Baconnier, P. et al. Self-aligning polar active matter. *arXiv:2403.10151* (2024).
46. Giomi, L., Kos, Ž., Ravnik, M. & Sengupta, A. Cross-talk between topological defects in different fields revealed by nematic microfluidics. *Proc. Natl. Acad. Sci. USA* **114**, E5771 (2017).
47. Pieranski, P., Hulin, J. & Godinho, M. H. Rheotropism of the dowser texture. *Eur. Phys. J. E* **40**, 619 (2017).
48. Emeršič, T. et al. Sculpting stable structures in pure liquids. *Sci. Adv.* **5**, eaav4283 (2019).
49. Čopar, S., Kos, Ž., Emeršič, T. & Tkalec, U. Microfluidic control over topological states in channel-confined nematic flows. *Nat. Commun.* **11**, 59 (2020).
50. Čopar, S. & Kos, Ž. Many-defect solutions in planar nematics: interactions, spiral textures and boundary conditions. *Soft Matter* **20**, 6894–6906 (2024).

## Methods
### Materials

4'-Pentyl-4-biphenylcarbonitrile (5CB) and E7 were obtained from Jiangsu Hecheng Advanced Materials. Silane dimethyl-octadecyl [3-(trimethoxysilyl) propyl] ammonium chloride (DMOAP; 42% weight in methanol) and glycerol were purchased from Sigma-Aldrich. Photoresist SPR220-4.5 was obtained from Microchem. Anhydrous ethanol was obtained from Decon Labs. Si wafers were purchased from Nova Electronic Materials and microscope glass slides were obtained from Fisher Scientific. All chemicals were used as received. Deionized water was purified to analytical grade using a Simplicity C9210 Milli-Q system.

### Fabrication of micropillar arrays

The micropillar array was fabricated by coating a Si wafer with photoresist, followed by UV exposure using a Karl Suss MJB-3 Contact Aligner to define the desired pattern. Subsequently, the patterned photoresist served as an etch mask in sculpting the surface of Si wafer via an ETC01 Technics reactive ion etching system. Removal of the photoresist was achieved through oxygen plasma exposure. The micropillar structures were verified with an FEI Quanta 200 scanning electron microscope (SEM).

### DMOAP functionalization of micropillar arrays

Microstructured Si wafers underwent triple 2-minute oxygen plasma exposures. Subsequently, they were immersed in a 1.5% v/v solution of DMOAP for 40 minutes, enabling the formation of a self-assembled monolayer of DMOAP to induce a homeotropic orientation of the LC on the Si surfaces. Afterward, the arrays were rinsed with water and ethanol to eliminate residual DMOAP and dried under a nitrogen stream.

### Characterization of LC orientation and topological defects in microstructure confinement

To investigate the orientational ordering and topological defects of the LC confined within the micropillar lattice structure, we infused 5CB into DMOAP-functionalized Si micropillar arrays. The infusion was achieved through a spin-coating process conducted at 800 rpm for one minute. Subsequent optical characterization of the LC-infused microstructure arrays was performed using an Olympus BX53M polarized light microscope operated in reflection mode. This analysis was conducted in both ambient air and submerged environments, providing insight into the LC orientational ordering across different mediums.

### Water droplet impact/removal-induced dipole alignment

To elucidate the response of micropillar array-confined LCs to the dynamics of water droplet impact and removal, a 10 μL water droplet was placed onto the LC-infused micropillar surface. Subsequently, 5 μL of water was precisely withdrawn from the droplet. The response of the dipoles was characterized using a polarized light microscope.

**Shear-induced dipole alignment**

Alignment of topological dipoles in micropillar array-confined LCs was achieved through water droplet sliding, water film shearing, or air bubble sliding underwater. Water droplet sliding involved placing a 10 µL water droplet on the LC surface, sliding it horizontally with a cantilever at a rate of approximately 1 mm/s, and carefully detaching the cantilever to preserve dipole orientation. For water film shearing, a 20 µL droplet was placed on the LC surface. Subsequently, a glass slide was positioned on the droplet, with a 2 mm spacer. Shear forces were applied by moving the top glass slide across the LC layer, inducing dipole reorientation parallel to the shear direction. We comment that the shear stress can be easily tuned by adjusting the moving velocity of the top glass slide, providing a simple and controllable method for dipole manipulation. For air bubble sliding, the confined LC was submerged underwater to a depth of 1 cm. Initial dipole alignment was achieved by directing water flow across the LC surface using underwater streaming, establishing a uniform dipole orientation. Subsequently, a 1−2 µL air bubble was introduced underwater onto the surface of the micropillar-confined LCs. A carefully attached cantilever at the bubble's top allowed controlled movement in a specified direction. The resulting alignment of topological dipoles was examined using a polarized light microscope.

**Creation of dipole domain walls**

To create domains of topological dipoles within micropillar array-confined LCs, we utilized droplet merging or underwater streaming. For the droplet merging approach, we initiated the experiment by positioning two identical 10 µL droplets on the LC surface confined by the micropillar array. The droplets were guided towards each other through manipulation using cantilevers until merging occurred. In the case of underwater streaming, we subjected the confined LC system to directed shear stress induced by water streaming onto the confined LC surface underwater, aligning the dipoles in the shear direction. The resulting domain patterns were observed under a polarized light microscope.

**Dipole manipulation via micro-cantilever**

Our methodology for manipulating dipole orientations within confined LCs leverages the application of local shear stress, achieved through precise underwater movements of a 50 µm-in-diameter micro-cantilever. This approach enabled us to generate targeted dipole orientation patterns, varying from individual dipole adjustments to the complex arrangement of entire dipole arrays into specified configurations.

**Fabrication of microfluidic chambers**

Microfluidic chambers were fabricated utilizing computer numerical control (CNC) milling, beginning with the design of the chamber geometry in Solidworks. The design was then transferred to Mastercam to generate toolpaths, facilitating the micro-milling of acrylic plastic bars into the desired structures using a 3-axis Roland Modela MDX-50 desktop CNC milling machine. Next, the chambers' inner surfaces were activated with oxygen plasma treatment to enhance bonding capabilities. The chamber bases were subsequently adhered to LC-infused micropillar array substrates using an adhesive, culminating in assembling the microfluidic device with an inner chamber height of 4 mm and an inlet/outlet diameter of 0.5 mm.

**Numerical analysis of LC ordering in micropillar arrays**

To investigate the LC ordering confined within micropillar array, we developed numerical sketches utilizing an analytical Ansatz. This Ansatz, constructed from complex analytical functions featuring poles and zeroes to represent defects with negative and positive winding numbers[50], offers an exact solution under the assumption of point-like pillars and pre-determined defect locations. This foundational model was then adapted to a 2D framework to approximate the boundary conditions of finite-sized pillars and facilitate defect relaxation to equilibrium positions. The model focused on minimizing the dimensionless free energy density ($F$), represented as:

$$F = s^4 - 2s_0^2 s^2 + \frac{1}{2}\xi^2 (\nabla s)^2 \tag{1}$$

where $s^2 = u^2 + v^2$ expresses the scalar order parameter $s$ with the diagonal and off-diagonal ($u, v$) components of the two-dimensional Q-tensor. The equilibrium order $s_0$ was arbitrarily set to 0.75. The nematic characteristic length $\xi$ was chosen as 1/20 of the lattice spacing. We comment here that these simulations primarily serve to sketch the director field for theoretical insights rather than to provide precise quantitative predictions, which would require taking into account the full 3D field and real material parameters. Accordingly, the energy scale and detailed 3D aspects of defects, as depicted in Fig. 1g, are considered less critical for the overarching theoretical framework presented.

**Numerical simulation of simplified XY model for dipole relaxation**

To model dipolar ordering at the simplest level, we performed simulations using a modified XY rotor model on a 2D triangular lattice of size 16 × 16, where each lattice site represents a pillar in the experimental system. The potential energy ($U$) of the system is defined as follows:

$$U = \Sigma_{<i,j>} (\theta_i - \theta_j)^2 - \Sigma_i A\cos(6\theta_i) \tag{2}$$

where $\theta_i$, measured with respect to the vertical axis, represents the angular position of the companion –1 defect adjacent to each pillar. The first term favors parallel alignment with nearest-neighbour interactions, while the second term favors alignment along multiples of 60°, along nearest-neighbour directions in the triangular lattice. The system is initialized with random values of $\theta_i$ between –π and π. Angular rotation of each dipole is modeled via overdamped dynamics with angular velocity proportional to the torque $-dU/d\theta_i$.

The ground state of this model system is a monodomain, but with the parameter $A > 0$ and fast quench via overdamped dynamics, it is trapped in metastable states. For a given quench rate, the resulting metastable states show cluster size that depends on $A$, with larger values of $A$ leading to smaller clusters. Supplementary Movie 2 demonstrates a qualitative match with experimental relaxation. In the simulation, dipole orientation is indicated by a white dot adjacent to each pillar, with similarly oriented dipole clusters indicated by pillar color. While this highly simplified model captures fundamental aspects of collective dipole ordering, it does not account for more complex topological phenomena, as discussed in the Results section.

**Acknowledgements**

This work is supported by NSF CMMI-2227991, the startup fund of The Ohio State University (OSU), OSU Materials Research Seed Grant Program, funded by the Center for Emergent Materials, an NSF-MRSEC, grant DMR-2011876, the Center for Exploration of Novel Complex Materials, and the Institute for Materials Research. S.Č. and U.T. acknowledge the support of the Slovenian Research and Innovation Agency (ARIS) through grant numbers P1-0055, P1-0099, J1-50006, J2-50092, and BI-US/24-26-087. R.L.B.S. acknowledges support from the United States-Israel Binational Science Foundation grant number 2022197.

**Author contributions**

U.I.K., B.C., S.Č., U.T., and X.W. conceived and designed the experiments, as well as lead the manuscript writing. U.T., and X.W. supervised the experiments. S.L. and Y.Y. fabricated the Si microstructure arrays, U.I.K., B.C., R.M., E.B., and Z.Y. conducted microstructure functionalization, polarized light microscopy characterization and data collection. S.Č. conducted the numerical modelling of the nematic LC director profile and accessible polar states. R.L.B.S. conducted numerical simulation of the simplified XY model of dipole relaxation to metastable states. All authors contributed to data interpretation, discussions, and manuscript preparation. U.I.K., B.C., and S.Č. contributed equally to this work.

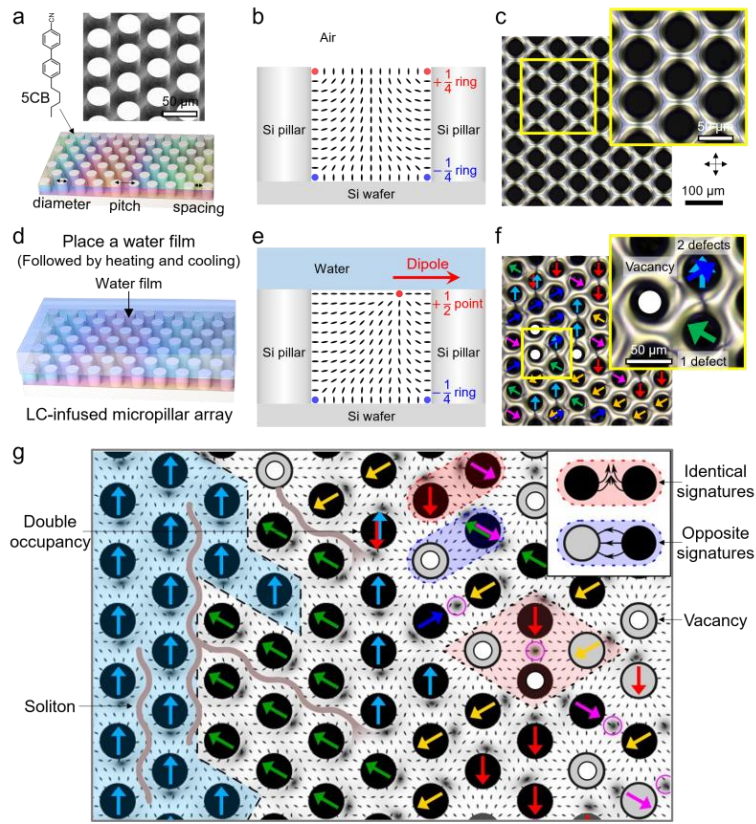

**Fig. 1 | Elastic dipoles in micropillar array-confined LCs. a**, Molecular structure of 5CB and schematic and representative SEM image of Si micropillar arrays. **b-f**, Polarized light micrographs of micropillar array-confined LCs and the director structure in the vertical cross-section for (**b**,**c**) air (homeotropic) and (**d-f**) water (planar) on the top interface. The air interface displays surface Saturn ring defects, and the water interface reveals displaced surface boojums with a winding number of −1 in the top view and +1/2 in the depicted vertical cross-section. Consequently, adjacent boojums to the pillars form oriented elastic dipoles (indicated by red arrow in **e**), with a few defects displaced, leading to some pillars having two or no accompanying defects (see inset). Crossed, double-headed arrows indicate the direction of the crossed polarizers. **g**, A representative 2D model of micropillar array-confined LCs. The micropillar color corresponds to their signature, where opposite signatures allow for direct streamlines, as shown in the inset and two encircled pillar pairs. The left half exhibits large domains of aligned dipoles (shaded in light blue), and most pillars share the same signature, along with a few highlighted solitons, which are consistent with experimental observations. The right half of the model demonstrates greater polarity variation, featuring centered defects (encircled in magenta), a phenomenon that is topologically possible and energetically favorable but not observed in experiments. The elastic dipole orientation is indicated by the color of the histogram as defined by Supplementary Fig. 2a. White disks indicate the vacancies (micropillars without an associated defect).

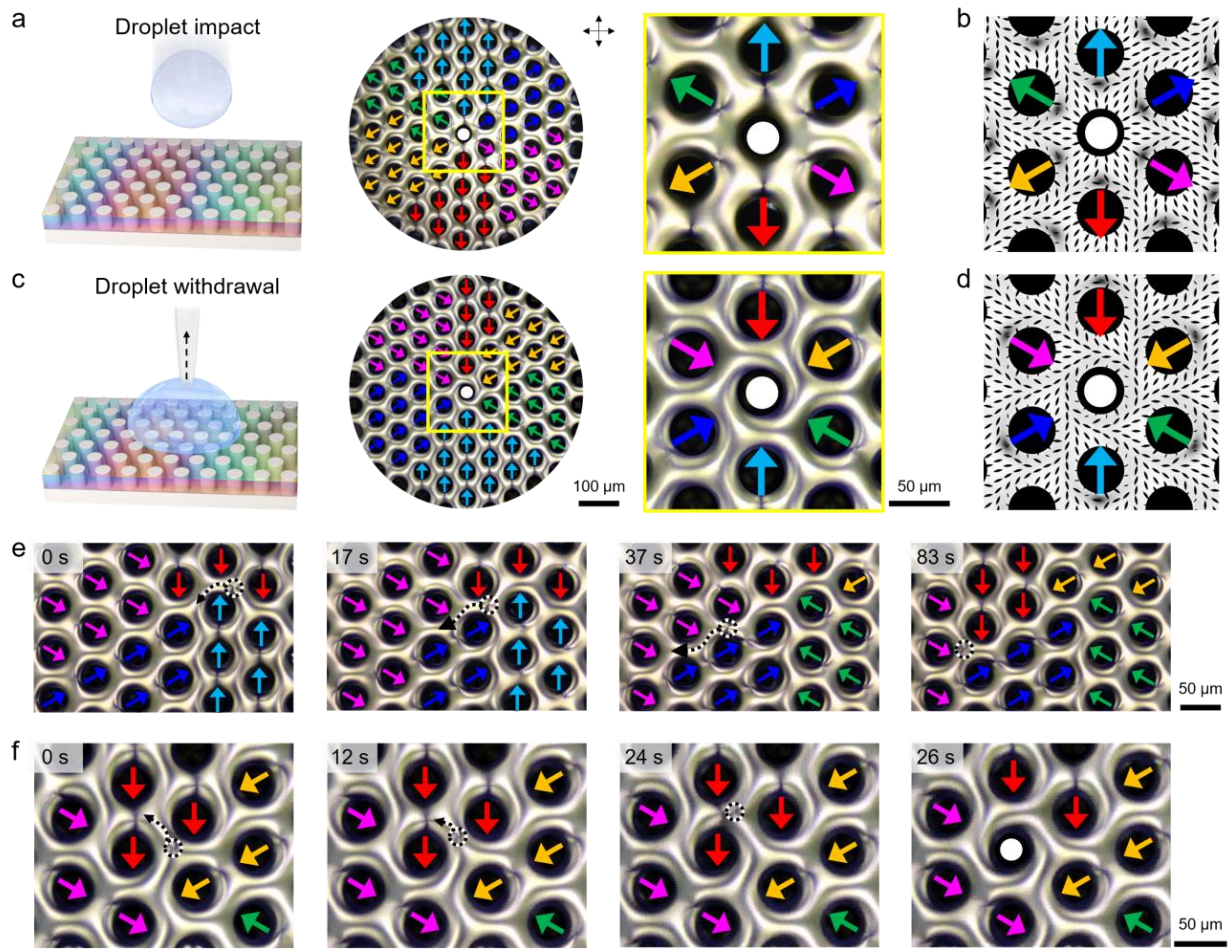

**Fig. 2 | Elastic dipole configuration induced by droplet placement and removal. a-d**, Schematic illustration, polarized light micrograph and numerical model depicting dipole orientations caused by (**a**,**b**) placement and (**c**,**d**) removal of a water droplet. A fallen droplet induces an outward radial water flow, resulting in the central micropillar being defect-free, effectively behaving as a +1 winding number defect within the discrete dipolar lattice. The central micropillar, featuring an opposite signature, facilitates direct streamlines towards neighbouring micropillars. Conversely, suction-induced inward flow creates an inward pointing defect, causing the central micropillar to display a spiral director due to the initially shared signature among all micropillars, requiring a π turn between adjacent micropillars. **e**,**f**, Polarized light micrographs showing that an unbound +1 point defect can either (**e**) traverse through the lattice, reorienting nearby dipoles (see Supplementary Movie 3), or (**f**) annihilate with a −1 point defect, resulting in the formation of a vacancy that maintains its original signature (see Supplementary Movie 4). Crossed, double-headed arrows indicate the direction of the crossed polarizers. White disks indicate the vacancies.

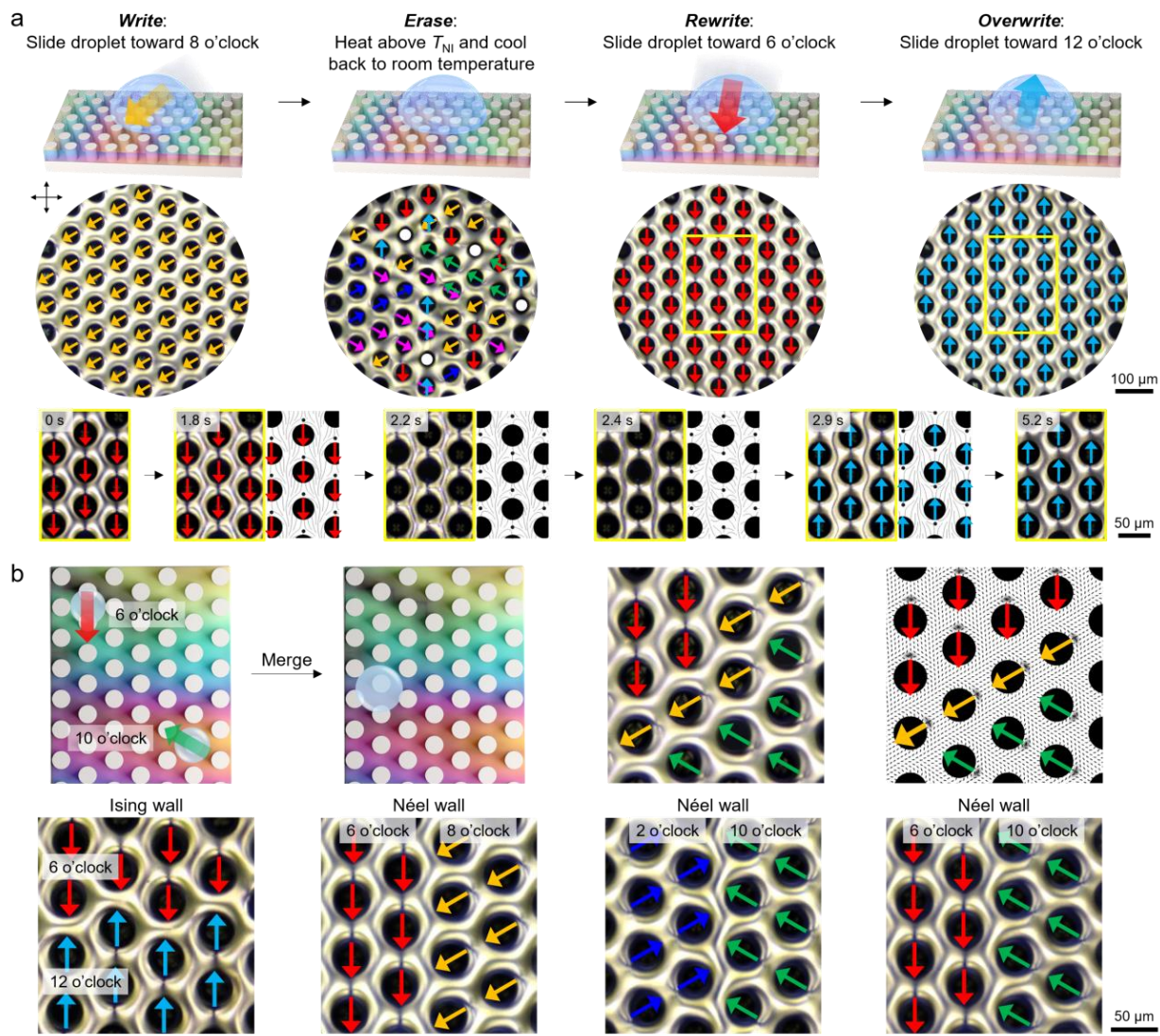

**Fig. 3 | Dipole reorientation induced by interfacial droplet motion. a**, Schematic and polarized light micrographs demonstrating that the dipole orientation can be written by sliding droplets, erased by temperature cycling through the isotropic phase, and rewritten or overwritten by sliding droplets (see Supplementary Movie 5). The bottom row shows polarized light micrographs and numerical analysis illustrating dipole reorientation from 6 o'clock to 12 o'clock. Supplementary Note 3 and Fig. 9 elucidate the dependence of threshold shear stress for dipole reorientation on the applied shear stress and initial dipole orientation. Crossed, double-headed arrows indicate the direction of the crossed polarizers. White disks indicate the vacancies. **b**, Schematic illustration, polarized light micrographs and numerical analysis demonstrating the formation of domain walls through droplet collisions.

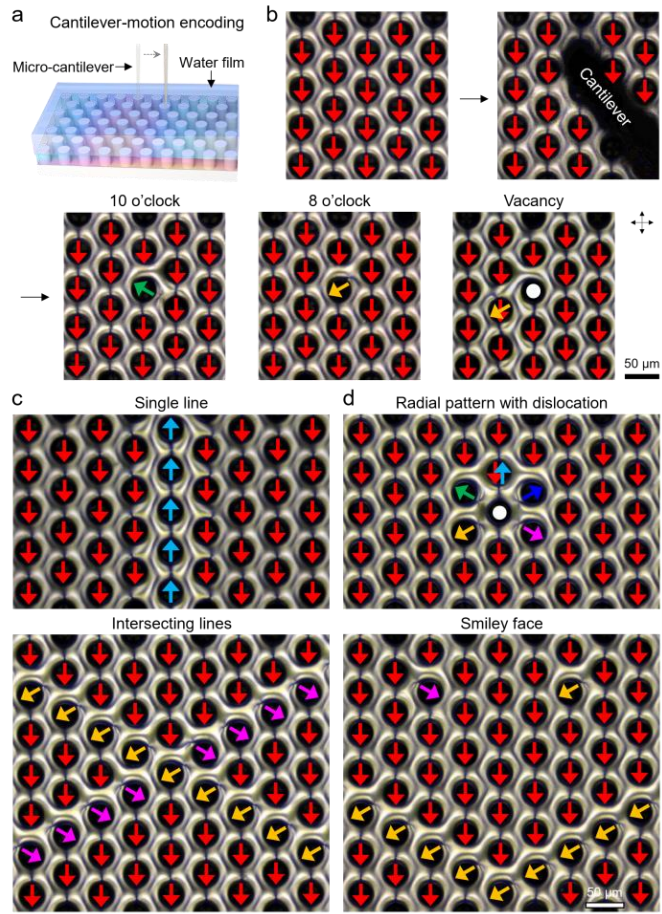

**Fig. 4 | Micro-cantilever encoding of dipole orientation. a**, Scheme showing manipulation of dipoles through micro-cantilever-induced shear. **b-d**, Polarized light micrographs demonstrating (**b**) the manipulation of a single dipole (see Supplementary Movie 7), and the creation of (**c**) single and intersecting lines and (**d**) complex patterns including radial pattern and a 'smiley face.' Dipoles are preset to 6 o'clock. Crossed, double-headed arrows indicate the direction of the crossed polarizers. White disks indicate the vacancies.

18